\documentclass[10pt]{fsttcs}         % LaTeX 2e 

\usepackage{color}
\usepackage{amsmath} \usepackage{amscd} \usepackage{amssymb} \usepackage{amsfonts} \usepackage{array}
\usepackage{graphics}
\usepackage[all]{xy}

\usepackage[pdftex]{graphicx}

\newtheorem{assumption}{Assumption}

\definecolor{blue}{rgb}{0,0,1}
\definecolor{red}{rgb}{1,0,0}

\newcommand{\mi}[1]{{\mathit{#1}}}
\newcommand{\msf}[1]{{\mathsf{#1}}}
\newcommand{\mbf}[1]{{\mathbf{#1}}}

\begin{document}

\title{Automatic analysis of distance bounding protocols}
%~\footnote{Accepted for presentation at an informal workshop in August 2009 (FCS'09).}

\begin{flushright}
\large{\textbf{ \sf Sreekanth Malladi\footnote{Dakota State University, USA, 	Email: \texttt{malladis@pluto.dsu.edu}}, Bezawada Bruhadeshwar$^{\dagger}$, Kishore Kothapalli\footnote{International Institute of Information Technology, India, Email: 		\texttt{\{bbruhadeshwar,kkothapalli\}@iiit.ac.in}
} }}
\end{flushright}

%\author{  					Sreekanth Malladi \\
%          Dakota State University\\
%          Madison, SD - 57042\\
%          \texttt{malladis@pluto.dsu.edu}           
%}

%\author{
%			Bezawada Bruhadeshwar, Kishore Kothapalli \\
%	  	International Institute of Information Technology \\	
%			Hyderabad, India \\
%			\texttt{\{bbruhadeshwar,kkothapalli\}@mail.iiit.ac.in}
%}	

\thispagestyle{empty}

\begin{abstract}

Distance bounding protocols are used by nodes in wireless networks for the crucial purpose of estimating
their distances to other nodes. This typically involves sending a request by one node to another node, receiving a response, and then calculating an upper bound on the distance by multiplying the round-trip time with the velocity of the signal. However, dishonest nodes in the network can turn the calculations
both illegitimate and inaccurate when they participate in protocol executions. Therefore,
it is important to analyze protocols for the possibility of such violations. Past efforts to analyze distance
bounding protocols have only been manual. However, automated approaches are important
since they are quite likely to find flaws that manual approaches cannot, as witnessed many times in the literature of key establishment protocols.

~~~~~~~~~~~~~~~~In this paper, we use the constraint solver tool to automatically analyze distance bounding protocols: We first formulate a new trace property called ~~{\em Secure Distance Bounding} (SDB) ~~that protocol executions must satisfy. We then classify the scenarios in which these protocols can operate considering the (dis)honesty of nodes and location of the attacker in the network. Finally, we extend the constraint solver tool so that it can be used to test protocols for violations of SDB in those scenarios and illustrate our technique on several  examples that include new attacks on published protocols. We also hosted an on-line demo for the reader to check out our implementation.

\end{abstract}

%\noindent %\textsc{Keywords.} \\  \noindent {\small	Wireless networks, Sensor networks, Localization, Distance bounding, Formal methods, Constraint solving, Cryptographic protocols, Timed analysis.}

\section{Introduction}\label{s.intro}

A ~~{\em distance bounding (DB) protocol} ~~is used by a ~~``verifier" ~~node in wireless networks to calculate an upper bound on the distance to a ~~``prover" ~~node in the network. Distance bounding helps in crucial applications such as localization, location discovery and time synchronization. Hence, the security of DB protocols is an important and critical problem.

\begin{flushleft}
\begin{figure}[h]
\scalebox{1.0}
{\includegraphics*[viewport=58 600 540 730]{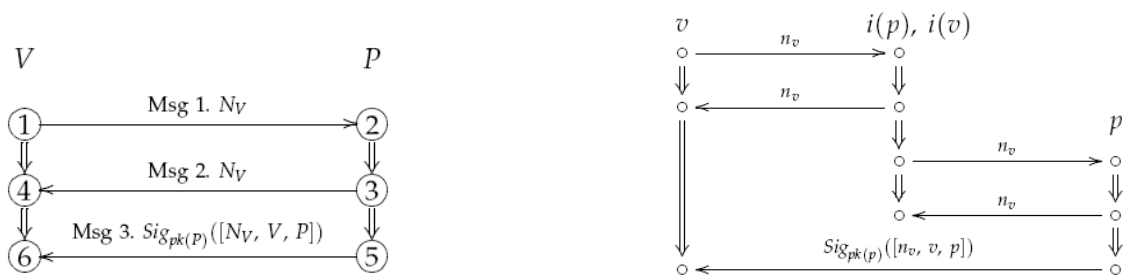}}
\caption{~~(a)~~~Extended Echo protocol {\bf P1}~~~~~~~~(b)~~~Man-in-the-Middle Attack on {\bf P1}}\label{f.P1-Protocol-Attack}
\end{figure}
\end{flushleft}

As an example of a DB protocol, consider a simple extension of the Echo protocol (Fig. \ref{f.P1-Protocol-Attack}.a)  presented in \cite{SSW03}. In the figure, ~~$\mi\mbf{V}$ ~~is the verifier, ~~${P}$ ~~is the prover; ~~$N_V$ ~~is a nonce; ~~$\mi{Sig}_{\mi{pk}(P)}([N_V,~V,~P])$ ~~is the signature of ~~$P$ ~~to be verified with it's public-key, denoted ~~$\mi{pk}(P)$. Let ~~$t_i$ ~~be the time on the clock when event ~~$i$ ~~occurs. Then, ~~$V$ ~~can calculate the bound ~~`$d$' ~~on the distance to ~~$P$ ~~as: ~~$d = \frac{(t_4 - t_1)~-~(t_3 - t_2)}{2} ~~\times ~~s$, ~~where ~~`$s$' ~~is the speed of the signal.

In the presence of attackers, DB protocols can fail  to achieve their main goal of establishing a valid distance bound. For instance, the above protocol has a possible attack wherein an attacker ~~${i}$ ~~plays Man-In-The-Middle and succeeds in showing ~~${p}$ ~~as being closer to ~~${v}$\footnote{We use lower case for ~~${v}$ ~~and ~~${p}$ ~~now since we are referring to the protocol execution.}  ~~than it really is (Fig. \ref{f.P1-Protocol-Attack}.b).

Analysis of DB protocols involves examining whether it is possible to make a party appear closer than it really is, to an honest verifier. The problem is different and difficult compared to standard Dolev-Yao analysis of protocols that only consider whether an attacker can generate messages required to violate some security property. Here, we need to factor in the time required for message generation as well, which can vary based on the input size and cryptographic parameters.  Automated analysis is much desired, given the problems and distrust in manual analysis of protocols that have been reported in literature \cite{LoweBF96}. There have been numerous instances when automated techniques found attacks on protocols that manual, hand-based techniques could not (e.g. \cite{Lowe96,Mea96,MS01}). 

\paragraph*{Past work.} The few published efforts to analyze DB protocols have been largely incomplete:  The classical work of Brands and Chaum \cite{BC95} 					is mostly informal and specific to the protocols 						introduced in that paper.  Sastry et al. \cite{SSW03} show that in their  							``Echo" protocol, the prover cannot respond before 					receiving the verifier's nonce but the protocol is 					used only for ``in-range" verification and also too 				simple without any authentication.  Meadows et al. \cite{MPPCS07} give a method to 							analyze both distance bounding and authentication 					aspects, but the method like the previous two, is 					manual, not automated.

\paragraph*{Our contribution.}
To address these concerns, we will show a method to automatically analyze DB protocols using the constraint solving technique of Millen-Shmatikov. Our method is based on formal modeling of timed protocols and distance bounding properties. Further, it is fully automated with minor changes to the existing constraint solver\footnote{on-line demo at ~~ \texttt{http://homepages.dsu.edu/malladis/research/ConSolv/Webpage/}}. Some highlights of our contribution are:

\begin{enumerate}

\item Like many past strand space extensions, our formal modeling and framework 					give a simple, clean and useful geometric flavor to the 				study of DB protocols that could be used or extended to 				many other studies such as  localization algorithms;

\item Some properties we prove about DB protocols allow the use 			of conventional Dolev-Yao style analysis, completely 						eliminating the need to consider the more 											complicated timing aspects. This is useful when it is 					difficult to extend existing methods for conventional key 			establishment protocols to analyze or verify DB 								protocols (e.g. {\sf ProVerif} \cite{BlanchetOakland06}).

\end{enumerate}

\paragraph*{Organization.} We will first develop a timed protocol model extending strand spaces in Section~\ref{s.prot-model}. We will then explain how constraint solving can be used to generate timed protocol executions in Section~\ref{s.ConSolv}. We will formalize secure distance bounding and explain our technique to detect violations for it in Section~\ref{s.distBound}. We will identify the scenarios under which DB protocols need to be analyzed in Section~\ref{s.scenarios}. We will illustrate our analysis approach on some examples in Section~\ref{s.testing}, and conclude with a discussion of future and related works.

\section{Protocol model - Timed strand spaces}\label{s.prot-model}

Our protocol model is based on the strand space model of \cite{THG98} extended with the introduction of a new field, ~~``time" ~~for labels on nodes. This field is used to represent the current time on the clock at the node for an agent.

\begin{definition}~~{\em\bf [Node]}\label{d.node} ~~A ~~\emph{node} ~~is a 3-tuple with fields ~~\emph{time, ~~sign}, ~~and ~~{\em term}. ~~Time is the current time on the clock, sign can be ~~$+$ ~~or ~~$-$ ~~denoting ~~``send" ~~and ~~``receive'' ~~respectively and term to be defined next.

\end{definition}

We will describe how to populate the times on nodes partly in this section and partly in the next section. We consider protocols in which messages are constructed using a free term algebra:

\begin{definition}~~{\em\bf [Term]}\label{d.term-alg} ~~A ~~\emph{term} ~~is one of the following: 	 ~~\emph{Variable} ~~(can be of types ~~$\mi{Agent}$, ~~$\mi{Nonce}$ ~~etc.);
	 ~~\emph{Constant} ~~(numbers ~~1,~2,~$\ldots$; name of the attacker ~~$\epsilon$ ~~etc.);
	 ~~\emph{Atom};
	 ~~\emph{Pair} ~~denoted ~~$[t_1,~t_2]$ ~~if ~~$t_1$ ~~and ~~$t_2$ ~~are terms;
	 ~~\emph{Public-Key} ~~denoted ~~$\mi{pk}(A)$ ~~with ~~$A$ ~~of type ~~$\mi{Agent}$;
	 ~~\emph{Shared-Key} ~~denoted ~~$\mi{sh}(A,B)$ ~~with ~~$A$ ~~and ~~$B$ ~~of type ~~$\mi{Agent}$;
	 ~~\emph{Asymmetric encryption} ~~denoted ~~$[t]^{\to}_k$ ~~where ~~$t$ ~~and ~~$k$ ~~are terms;
	 ~~\emph{Symmetric encryption} ~~denoted ~~$[t]^{\leftrightarrow}_k$ ~~where ~~$t$ ~~and ~~$k$ ~~are terms;
	 ~~\emph{Hash} ~~denoted ~~$h(t)$ ~~where ~~$t$ ~~is a term;
	 ~~\emph{Signature} ~~of a term ~~$t$ ~~denoted ~~$\mi{Sig}_{\mi{pk}(A)}(t)$ ~~to be validated using  ~~$\mi{pk}(A)$.

\end{definition}

%A ``parametric term" is any term with only variables in it. 
A ~~``ground" ~~term is any term with no variables in it. 
We will drop the superscript ~~$\to$ ~~or ~~$\leftrightarrow$ ~~if the mode of encryption is contextually either obvious or irrelevant. 

\begin{definition}~~{\em\bf [Subterm]} \label{d.subterm} ~~Term ~~$t$ ~~is a ~~\emph{subterm} ~~of ~~$t'$ ~~(i.e. ~~$t \sqsubset t')$ ~~if ~~$t = t'$, ~~or ~~if $t' = [t_1, t_2]$ ~~with ~~$t \sqsubset t_1 \vee t \sqsubset t_2$, ~~or ~~if $t' = [t^{''}]_{k'}$ ~~with ~~$t \sqsubset t^{''}$, ~~or if ~~$t' = h(t^{''})$ ~~with ~~$t \sqsubset t^{''}$, ~~or if ~~$t' = \mi{Sig}_{\mi{pk(A)}}(t^{''})$ ~~with ~~$t \sqsubset t^{''}$. ~~Term ~~$t$ ~~is a ~~\emph{proper subterm} ~~of ~~$t'$ ~~if ~~$(t \sqsubset t')\wedge(t \neq t')$.

\end{definition}

Strands capture roles of a protocol.

\begin{definition}~~{\em\bf [Strand]} ~~A ~~\emph{strand} ~~is a sequence of nodes. For instance ~~$s = \langle n_1, \ldots , n_m \rangle$ ~~is a strand with $m$ nodes. Nodes in a strand are related by the edge ~~$\Rightarrow$ ~~defined such that if ~~$n_i$ ~~and ~~$n_{i+1}$ ~~belong to the same strand, then  ~~$n_i \Rightarrow n_{i+1}$. ~~A ~~\emph{parametric strand} ~~is a strand with no atoms in the terms on its nodes.

\end{definition}

Protocol roles are modeled as partially instantiated parametric strands that we name ~~\emph{semi-strands} ~~where messages contain variables and atoms depending on the knowledge of agents concerning message subparts. For instance, the verifier strand of the protocol presented in the Introduction is represented as 

\[ 
		\langle~~+[0,~n_v], ~~-[T_4,~n_v], 
				~~-[T_6,~\mi{Sig}_{\mi{pk}(P)}([n_v,~v,~P])]~~\rangle 
\]

Notice that the first node starts at time ~~`0' ~~which is not a universal ~~`0' ~~but a local start time for the agent who dons this strand. Also notice that the times on the other two nodes ~~$T_4$ ~~and ~~$T_6$ ~~are not fixed. The rationale for this is to be explained shortly.

 A set of semi-strands is called a ~~\emph{semi-bundle}. ~~We will say that term $t$ belongs to a semi-bundle ~~$S$ ~~(i.e. ~~$t \in S$) ~~if ~~$(t = \mathrm{term}(n))$ ~~for some ~~$(n \in s)$ ~~and ~~$(s \in S)$. 

A ~~\emph{bundle} ~~is a possible protocol execution obtained by consistently instantiating all the variables in the semi-bundle and using ~~$\to$ ~~edges between nodes on different strands.

\begin{definition} ~~{\em\bf [Bundle]} ~~A ~~\emph{bundle} ~~is a collection of strands and an acyclic digraph defined on a mapping of nodes to edges ~~$\to$ ~~and ~~$\Rightarrow$ ~~such that if node ~~$n_i$ ~~sends a message that ~~$n_j$ ~~receives, then ~~$n_i, ~~n_j$ ~~are related by the edge ~~$\to$ ~~(denoted ~~$n_i \to n_j$). ~~Further, if there is a node ~~$n$ ~~in the bundle that receives a term ~~$t$, ~~then there is another node ~~$m$ ~~in the bundle, that sends ~~$t$ ~~such that ~~$m \to n$.

\end{definition}

Note that this bundle is a 3-dimensional graph with strands located vertically anywhere in the cube. Nodes in a bundle are also related by precedence relation denoted ~~$\preceq$ ~~which is a partial order:

\begin{definition} ~~{\em\bf [Precedes]} ~~The relation ~~$\preceq$ ~~is defined such that if nodes ~~$n_i, ~~n_j$ ~~exist in a bundle ~~$\mathcal{C}$, ~~then ~~$n_i \preceq n_j$ ~~if they are on the same strand with ~~$i \leq j$; ~~further, ~~$n_i \prec n_j$ ~~if ~~$n_i \to n_j$.

\end{definition}

We will use ~~$\preceq$ ~~on stand-alone strands in semi-bundles as well: Let ~~$s$ ~~be a strand in a semi-bundle ~~$S$. ~~Then, ~~$(\forall n_i, n_j \in s)(s \in S)(i \leq j \Rightarrow n_i \preceq n_j)$.

We do not include the notion of penetrator strands as in the classical strand spaces formalism of \cite{THG98}. Rather, we consider a single penetrator also modeled as a single strand that captures all the ~~``penetrator actions'' ~~in the bundle defined as below:

\begin{definition} ~~{\em\bf [Penetrator action]}\label{d.pen-action} ~~A ~~\emph{penetrator action} ~~is a sequence of edges ~~$t_1 \to t_2 \Rightarrow t_3 \to t_4$ ~~where ~~$t_2 \Rightarrow t_3$ ~~is an edge on the penetrator strand.

\end{definition}

The idea is that the single ~~$\Rightarrow$ ~~edge in a penetrator action represents all the penetrator strands in the classical model of \cite{THG98} to generate the term to be sent. Multiple penetrators could be added in the 3-dimensional cube if desired, although we only consider a single ``Machiavellian" attacker with full control of the network in the spirit of \cite{SM00}\footnote{This might be unrealistic in wireless networks, but the stronger model allows us to find all attacks including those under weaker attackers.}. 

Next we define the ~~``elapsed time" ~~between any two nodes ~~$n_i, ~~n_j$ ~~in a bundle ~~$C$ ~~with ~~$n_i \preceq n_j$ ~~using the notion of ~~\emph{weights} ~~and ~~\emph{paths}:

\begin{definition} ~~{\em\bf [Weight or Elapsed time]} ~~The ~~\emph{weight} ~~of an edge is the (absolute) difference in times between the nodes that are connected by the edge. A ~~\emph{path} ~~is a sequence of nodes such that every node in the sequence is related to the subsequent node by a ~~$\to$ ~~or a ~~$\Rightarrow$. ~~The weight of a path is the sum of the weights of all the edges in the path.	

\end{definition}

We will denote the path between ~~$n_i$ ~~and ~~$n_j$ ~~as ~~$(n_i,~n_j)$ ~~when there is only one route between ~~$n_i$, ~~and ~~$n_j$.

The weight of a ~~$\pm t \Rightarrow +t'$ ~~edge should be preset and constant for each semi-strand. In the case of penetrator strand, those weights should be calculated using penetrator actions required to generate the ~~$+$ ~~node. On the other hand, the weight of a ~~$\pm t \Rightarrow -t'$ ~~edge cannot be fixed since an agent can only know the length of time after which it sends a message, but cannot always predict when it might receive a message from another agent, accurately.

Weights of ~~$\to$ ~~edges indicate the  time of traversal for  messages which depends on the message length, distance and the velocity of the signal. We assume that there is an appropriate formula for an environment to calculate the weight of these edges, using those parameters.

\begin{definition} ~~{\em\bf [Relay, ~~Simple relay]} \label{d.relay} ~~A ~~\emph{relay} ~~is a penetrator action ~~$+t \to -t \Rightarrow +t \to -t$. ~~A ~~\emph{simple relay} ~~is a relay with the weight of the ~~$\Rightarrow$ ~~edge being zero. 
\end{definition}

We develop the notion of ~~``ideal" ~~and ~~``real" ~~bundles to distinguish protocol executions where the penetrator plays a passive role of merely observing message exchanges between agents with those where she plays an active role of faking and changing messages.

\begin{definition} ~~{\em\bf [Ideal and Real bundles]} \label{d.ideal-real} ~~An ~~{\em ideal bundle} ~~$B$ ~~for a protocol ~~$P$ ~~is a bundle formed from a semi-bundle ~~$S$ ~~with exactly one semi-strand per parametric strand of ~~$P$ ~~where every penetrator action is a simple relay ~~$+\sigma t \to -\sigma t \Rightarrow +\sigma t \to -\sigma t$ ~~for some substitution ~~$\sigma$ ~~such that ~~$(\forall s \in S)((\exists s' \in B)(s' = \sigma s))$. ~~A ~~{\em real bundle} ~~is any bundle from any other semi-bundle from ~~$P$.
\end{definition}

\section{Extending constraint solving to find elapsed time}\label{s.ConSolv}

We will now extend the constraint solving technique of \cite{MS01} to give a ``recipe" to produce the timed bundles  defined in Section~\ref{s.prot-model} including honest strands and the single penetrator strand with all the penetrator actions. 

The previous section only noted that weights of ~~$\pm \Rightarrow +$ ~~edges should be preset; this section will complete labeling of nodes since weights on ~~$\pm \Rightarrow -$ ~~edges are calculated dynamically setting the times on ~~`$-$' ~~nodes during protocol executions. The elapsed time between any two nodes in such bundles can then be calculated by summing up the weights on all the edges in the path between the nodes.

%\subsection{Constraint Solving - Background}

Constraint solving is a procedure to determine if a semi-bundle is completable to a bundle using a substitution to variables. A constraint sequence is first drawn from node interleavings of the semi-bundle indicating that ~~`$-$' ~~nodes should be derivable by the attacker with his actions and terms on all prior ~~`$+$' ~~nodes.

\begin{definition} ~~{\em\bf [Constraint sequence]}\label{d.conseq} ~~A ~~{\em constraint sequence} ~~$C = \langle ~\mathrm{term}(n_1)~:~T_1,~\ldots,$ $~\mathrm{term}(n_k)~:~T_k ~\rangle$ ~~is from a semi-bundle ~~$S$ ~~with ~~$k$ ~~`$-$' ~~nodes if ~~$(\forall ~n,~n')( ~((\mathrm{term}(n')~:~T \in C)$  $\wedge (\mathrm{term}(n) \in T)) \Rightarrow (n \preceq n') ~)$. ~~Further, if ~~$i < j$ ~~and ~~$n_i$, ~~$n_j$ ~~belong to the same strand, then ~~$n_i \preceq n_j$ ~~and ~~$(\forall i)(T_i \subseteq T_{i+1})$.

\end{definition}

We consider a set of attacker operators ~~$\Phi$ ~~and an infinite set of terms that can built using ~~$\Phi$ ~~on a finite set of terms ~~$T$, ~~denoted ~~$\mathcal{F}(T)$. ~~Although our techniques in this paper are largely independent of the kind of operators in ~~$\Phi$, ~~we will consider that they represent the standard Dolev-Yao attacker as defined in \cite{MS01}.

The possibility of forming bundles from a given semi-bundle can be determined by testing if constraint sequences from it are satisfiable:

\begin{definition}~~{\em\bf [Satisfiability,~~ Realizability]}\label{d.satisfiable} ~~A constraint is ~~$m : T$ ~~is ~~{\em satisfiable} ~~under a substitution ~~$\sigma$ ~~if ~~$\sigma m \in \mathcal{F}(T)$. ~~A constraint sequence is ~~{\em satisfiable} ~~with ~~$\sigma$, ~~denoted ~~$\sigma \vdash C$ ~~if ~~$(\forall m : T \in C)(\sigma m \in \mathcal{F}(\sigma T))$. ~~A ~~`$-$' ~~node is ~~{\em realizable} ~~if the corresponding constraint is satisfiable. ~~A semi-bundle is ~~{\em completable} ~~to a bundle if a constraint sequence from it is satisfiable. 

\end{definition}

Millen-Shmatikov have shown a constraint satisfaction procedure, denoted ~~{\bf P} ~~that is terminating, sound and complete wrt ~~$\Phi$ ~~and ~~$\mathcal{F}$. ~~{\bf P} ~~applies a set of symbolic reduction rules ~~$R$ ~~to each constraint, in order to reduce it them to ~~``simple constraints" ~~(with only a variable each on the left side). We provide both ~~{\bf P} ~~and ~~$R$ ~~in Appendix \ref{sa.consolv}.

We will consider that each reduction rule in ~~$R$ ~~corresponds to an attacker action and we will calculate the weights of ~~$\Rightarrow$ ~~edges of a bundle to be the sum of the times taken by each rule.

In Appendix \ref{sa.consolv} we also give an algorithm denoted ~~{\bf PB} ~~that produces timed bundles as defined in Section \ref{s.prot-model}, using ~~{\bf P} ~~to calculate the weights of ~~$\Rightarrow$ ~~edges. An example bundle generated for the Lowe's attack on the NSPK protocol \cite{LoweBF96} is also given in Appendix \ref{sa.alg-example}. Further, we show in Appendix \ref{sa.proofs}, Theorem \ref{ta.termination} that ~~{\bf PB} ~~terminates and is sound and complete.

\section{Analyzing DB protocols}\label{s.distBound}

We will now formalize secure distance bounding using the concept of ideal and real bundles defined in Section  \ref{s.prot-model}.

\subsection{Formalizing Secure Distance Bounding}

A DB protocol is used by a verifier ~~$v$ ~~to establish an upper  bound on the distance to a prover ~~$p$. ~~Ideally, if the following assumptions hold: ~~~~{\bf (a)} ~~The positions of ~~$v$ ~~and ~~$p$ ~~are fixed, ~~{\bf (b)} ~~The intervals between creating and sending messages are fixed, ~~{\bf (c)} ~~$v$, ~~$p$ ~~are honest and 
~~{\bf (d)} ~~There is no attacker; ~~~then there indeed exists an upper bound on the distance that can be calculated by calculating the elapsed time between two nodes ~~$\mi{Request}$ ~~and ~~$\mi{Response}$ ~~on ~~$v$ ~~with ~~$\mi{Request}$ ~~a send node, ~~$\mi{Response}$ ~~a receive node and ~~$\mi{Request} \prec \mi{Response}$ ~~as explained in Section \ref{s.prot-model}.

We will call the nodes between ~~$\mi{Request}$ ~~and ~~$\mi{Response}$ ~~in the verifier strand of a DB protocol as the ~~``DB part" ~~and the other nodes as the ~~``authentication part". ~~Further, we will use the term ~~``Time of Flight" ~~or its abbreviation as ~~ToF ~~to refer to the elapsed time between ~~$\mi{Request}$ ~~and ~~$\mi{Response}$. 

Now the upper bound that is calculated by ~~$v$ ~~can be lowered compared to the one obtained under ideal conditions, if ~~~{\bf (a)} ~~the ToF between its ~~$\mi{Request}$ ~~and ~~$\mi{Response}$ ~~is 				lowered ~~\emph{and} ~~~{\bf (b)} ~~if $v$ is sent all the messages in the protocol that it 					expects to receive from ~~$p$.

This is the main insight in defining secure distance bounding as a trace property: We first calculate the ~~ToF ~~under ideal conditions and check whether a ~~``real" ~~execution of the protocol in the presence of the penetrator can result in a calculation of ToF that is lower than the ideal value. Note that we assume  weights of ~~$\pm m \Rightarrow +m'$ ~~edges are set for strands in semi-bundles, by following the same measures to calculate time taken for message construction outlined in Section~\ref{s.ConSolv}.

\begin{definition}~~{\em\bf [Secure Distance Bounding (SDB)]}\label{d.SDB} ~~Let ~~$t$ ~~and ~~$t'$ ~~be the elapsed times in the verifier strand  of an ideal and real bundle ~~($B$ ~~and ~~$B'$) ~~respectively from a semi-bundle ~~$S$, ~~between the ~~$\mi{Request}$ ~~and ~~$\mi{Response}$ ~~nodes. Then  ~~\emph{Secure Distance Bounding (SDB)} ~~is satisfied in ~~$B'$, ~~whenever ~~$t < t'$. ~~Conversely, ~~SDB ~~is violated in ~~$B'$ ~~if ~~$t > t'$.

\end{definition}

This definition is dependent on what we consider an ideal bundle to be. In Section~\ref{s.prot-model}, we defined it to be one with no penetrator actions, but when the penetrator is further from ~~$v$ than $p$ is, we would need to make the bundle between the penetrator and $v$ as the ideal. More on this is explained in Section~\ref{ss.loc-scenarios}.

\section{Protocol execution scenarios}\label{s.scenarios}

Before explaining our technique to test protocols for violations of SDB, we will consider the scenarios under which a DB protocol can operate.

\subsection{Scenarios based on honesty of the prover}\label{ss.hon-scenarios}

We first consider scenarios in which the prover is honest  or dishonest. 

\paragraph*{Scenario A (honest prover).} With the verifier, honest prover and an attacker, this scenario captures MITM/Mafia attacks \cite{Desmedt88}. The attack described in the Introduction is one such attack.

\begin{center}
\begin{figure}[h]
\centering
\scalebox{1.0}
{\includegraphics*[viewport=70 600 540 730]{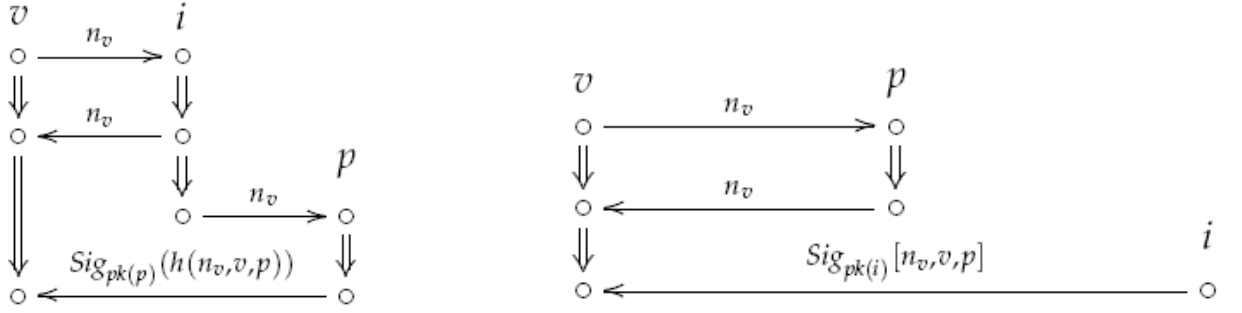}}
\caption{(a) Scenario B --- colluding attacker~~~~~~(b) Scenario 2 --- further attacker}\label{f.scenarios}
\end{figure}
\end{center}

\paragraph*{Scenario B (dishonest, colluding prover).} With the verifier, dishonest prover and attacker, this scenario captures terrorist/collusion attacks \cite{Desmedt88}. Here, the prover colludes with an attacker who is presumably closer to the verifier, by passing some or all of its information including secret keys and messages (partial or full collusion). The protocol in Section ~\ref{s.intro} is vulnerable to such an attack (Fig. \ref{f.scenarios}.a). 

%Notice that $p$ does not send the nonce $n_v$ back to $i$, since that is not required when they both collude.

\subsection{Scenarios based on location of attacker}\label{ss.loc-scenarios}

Independent of the honesty of agents, we should also categorize protocol execution scenarios based on the location of the attacker in the network with respect to the verifier and the prover.

\paragraph*{Scenario 1 (closer attacker).} Attacker ~~$i$ ~~physically closer to the verifier ~~$v$ ~~than the prover ~~$p$ ~~is. The first attack on ~~{\bf P1} ~~described previously is an example for this scenario. 

In this situation, we can show that ~~~~{\bf (a)} ~~~if an attacker can generate all the messages expected by the verifier from the ~~$\mi{Request}$ to the ~~$\mi{Response}$ ~~without those messages being sent by the prover ~~{\em and} ~~~~{\bf (b)} ~~~if all other messages expected by the verifier can also be generated by the attacker (with or without those messages emanating from the prover), ~~~then SDB is violated:

\begin{theorem}\label{t.closer} Suppose ~~$t_0,~t_1,~ \ldots,~t_m$ ~~are terms on ~~$m$ ~~nodes on the  verifier strand ~~$v$ ~~with time of flight measured in between ~~$t_0$ ~~and ~~$t_m$. ~~Then, there exists a bundle with a violation of SDB if 

\begin{itemize}
		\item the constraints ~~$\langle ~t_1 ~: ~T_0, ~\ldots ~,~t_m : T_m ~\rangle$ are satisfiable where for ~~$i = 0$ to $m$, 							~~every ~~$t \in T_i$ ~~either belongs to ~~$T_0$ ~~or a ~~$+$ 							~~node on ~~$v$ ~~and every ~~$t_i$ ~~is a term on a ~~$-$ ~~node on 					 ~~$v$;
		\item all other ~~$-$ ~~nodes in ~~$v$ ~~are realizable.	
\end{itemize}

\end{theorem}

\begin{proof}~~Please see Appendix \ref{sa.proofs}, Theorem \ref{ta.closer}.

\end{proof}

\paragraph*{Scenario 2 (farther attacker).} Attacker ~~$i$ ~~is physically farther from ~~$v$ ~~than ~~$p$. ~~Here, ~~$i$ ~~tries to show itself closer to ~~$v$ ~~by using the responses from ~~$p$ ~~to ~~$v$ ~~in the DB part, and then inserts its own messages for the authentication part. ~~{\bf P1} ~~is vulnerable in this scenario as well (Fig. \ref{f.scenarios}.b).

This scenario is exactly opposite of Scenario~1: we just have to assume that the ideal bundle now is in between ~~$v$ ~~and ~~$i$ ~~instead of ~~$v$ ~~and ~~$p$. ~~We should then analyze protocols for potential executions with ~~$p$ ~~sending all the messages in the DB part and attacker sending the remaining messages. We prove this below:

\begin{theorem}\label{t.farther} Consider ~~$v,~p_1, ~p_2$ ~~where ~~$d(v,p_1) < d(v,p_2)$. ~~Let ~~$\langle ~t_0,~\ldots,~t_m ~\rangle$ ~~be nodes on ~~$v$ ~~between which time of flight is measured.
	Then, there is a violation of SDB if 
	
	\begin{itemize}
		\item the constraints ~~$\langle ~t_1~:~T_1, ~\ldots, ~t_m~:~T_m ~\rangle$ ~~are satisfiable where for all ~~$i = 1$ to $m$, 	~~every ~~$t_i$ ~~is unified with some ~~$t'_1 \in T_i$ ~~where ~~$t'_i$ ~~is a term on ~~$p_1$.
		\item All other ~~`$-$' ~~nodes of ~~$v$ ~~are realizable  						without unifying with any subterms of ~~$p_1$.
	\end{itemize}
	
\end{theorem}

\begin{proof}~~~Please see Appendix \ref{sa.proofs}, Theorem \ref{ta.farther}.
\end{proof}

%This testing can be simulated by removing the nodes in the DB part in $i$'s strand but retaining the corresponding nodes in $p$'s strand and vice-versa for the remaining messages.

\section{Implementation and Examples}\label{s.testing}

	We now present some example protocols and their analyses using our technique. We tested all the protocols in the Constraint Solver tool with the scenarios and results in Section~\ref{s.scenarios}. We hosted all the protocols and scenarios in our on-line demo which can be tested with the click of a button. Here, we will present only the most interesting attacks and at least one per type of scenario.

It is worth mentioning that we made a simple change to the solver: we restricted it to consider only those node interleavings wherein the ~~$\mi{Request}$ ~~and ~~$\mi{Response}$ ~~nodes in the verifier strand immediately follow each other. We show in Appendix \ref{sa.sound-complete} that this is required to ensure soundness and that it preserves completeness wrt Def~\ref{d.SDB}.

In all the protocols below, distance bound, ~~$d = \frac{\delta_1 ~- ~\delta_2}{2} \times s$ ~~verifier fixes ~~$\delta_2$ ~~as a constant for a given protocol. Further to save space, we simplified some bundles by removing simple and insignificant relays.

%%%%%%%%%%%%%%%%%%%%%%%%%%%%%%%%%%%%%%%%%%%%
%%    Protocol P2 - Analysis							%%
%%%%%%%%%%%%%%%%%%%%%%%%%%%%%%%%%%%%%%%%%%%%

\subsection{P2 - Brands and Chaum \cite{BC95}}\label{ss.P2}

\begin{flushleft}
\begin{figure}[h]
\scalebox{1.0}
{\includegraphics*[viewport=70 580 540 730]{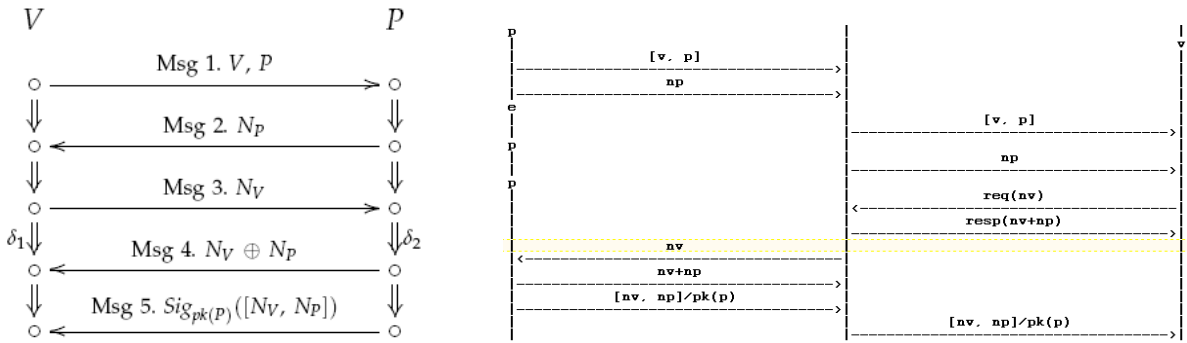}}
\caption{(a) Brands-Chaum protocol {\bf P2} ~~~~(b) Actual solver trace of the MITM Attack on {\bf P2} (NOTE: ~~{\tt nv+np = $[n_v]^{\leftrightarrow}_{n_p}$, [nv,np]/pk(p) = $\mi{Sig}_{\mi{pk}(p)}[n_v,~n_p]$}; ~~$\mi{Request}$ ~~and ~~$\mi{Response}$ ~~nodes were coded as ~~{\tt req(nv)} ~~and ~~{\tt resp(nv+np)} )}\label{f.P2-Protocol-Attack}
\end{figure}
\end{flushleft}

\noindent
The original Brands-Chaum protocol is a bit tricky with commit, rapid bit-level exchange and authentication/sign phases,  and \texttt{XOR} operator that is not modeled by the solver. Hence, we analyzed an approximate version (Fig. \ref{f.P2-Protocol-Attack}.a).

Notice that there is a pre-commitment of nonce ~~$N_P$ ~~by ~~$P$. ~~Brands and Chaum specify that messages 3 and 4 should be  bit-by-bit exchanges with the round-trip time calculated as the average of all the bit exchanges. Since the exchange is rapid and no other messages can interfere during the exchange, we felt it safe to  model the protocol with just one of those message exchanges. Also, ~~$N_V\oplus N_P$ ~~was modeled as ~~$[N_V]^{\leftrightarrow}_{N_P}$.

\paragraph*{Honest prover, Closer attacker.} Following our results in Section~\ref{s.scenarios}, we removed the nodes in the DB part in the prover strand and found an MITM attack on ~~{\bf P2} ~~which was similar to the MITM attack on ~~{\bf P1} ~~shown in the Introduction: Attacker simply sends all the messages except the signature to the verifier and later sends all of them to the prover. Finally, she relays the signature from the prover to the verifier. The solver found three different attack traces with three different node interleavings all essentially the same attack (Fig. \ref{f.P2-Protocol-Attack}.b).

The original Brands-Chaum protocol actually requires that the commitment ~~$N_P$ ~~be secretly exchanged between ~~$V$ ~~and ~~$P$. ~~With this requirement, the protocol forms a nice counter-example to Theorem~\ref{t.closer}: not all constraints corresponding to messages between Request (Msg~2) and Response (Msg~4) are satisfiable. When we made this change in the solver, it did not report an attack.

\paragraph*{Dishonest prover, closer attacker.} Obviously, revealing the nonce ~~$N_P$ ~~(the commitment) to the attacker before hand allowed the attack (partial collusion) and of course, full collusion worked too. In any case, Brands-Chaum seems stronger against collusion than ~~{\bf P1} ~~since it requires sharing of ~~$N_P$ ~~for the attack to succeed.

\paragraph*{Farther attacker.} This protocol also forms a nice example to test under Scenario~2. Assuming that the attacker is further away from the verifier, we followed our guidelines in Section \ref{s.scenarios} and removed the nodes in the DB part in one strand, while removing the signature (Msg~5) in another strand. The solver then output an attack where the agent whose DB part was removed looks closer than it is to the verifier (see Appendix \ref{sa.more-attacks}).
	
%%%%%%%%%%%%%%%%%%%%%%%%%%%%%%%%%%%%%%%%%%%%
%%    Protocol P3 - Analysis							%%
%%%%%%%%%%%%%%%%%%%%%%%%%%%%%%%%%%%%%%%%%%%%

\subsection{P3 - Meadows et al. \cite{MPPCS07}}\label{ss.P3}

\begin{flushleft}
\begin{figure}[h]
\scalebox{1.0}
{\includegraphics*[viewport=70 540 540 730]{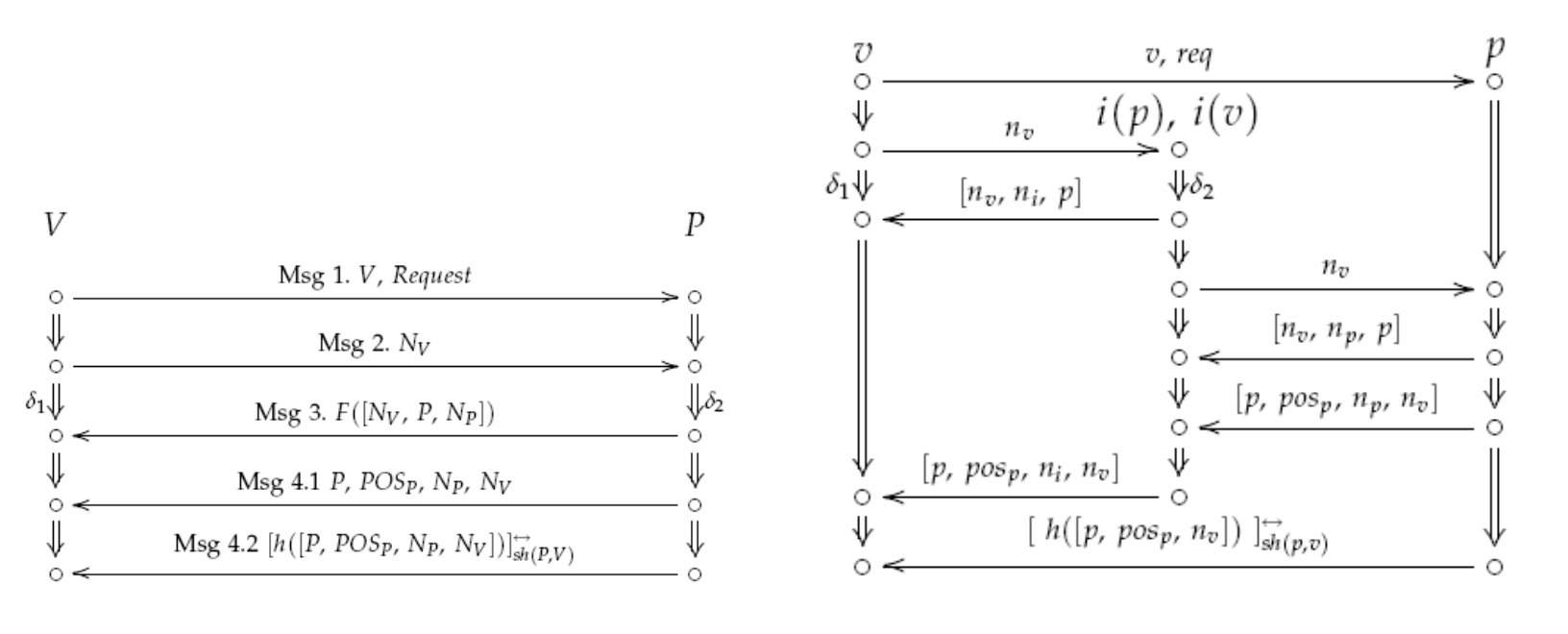}}
\caption{~~~~(a) ~~Meadows et al. protocol {\bf P3} ~~~~~~~~~~~~~~~~~~(b)~~ MITM Attack }\label{f.P3-Protocol-Attack}
\end{figure}
\end{flushleft}

{\bf P3} below was recently proposed in \cite{MPPCS07} (Fig. \ref{f.P3-Protocol-Attack}.a). 

\paragraph*{Honest prover, closer attacker.} ~~{\bf P3} ~~is actually quite similar to ~~{\bf P2} ~~and Brands-Chaum but with some crucial changes. Even without any commitment step, it was not vulnerable to the MITM attack that we presented in Section~\ref{ss.P2} even though the nonce ~~$N_P$ ~~is sent in plain in Msg~3 unlike Brands-Chaum that does not disclose it. This shows that sending ~~$N_P$ ~~\underline{before} ~~the ~~$\mi{Response}$ (Msg~3) ~~was the fatal mistake in ~~{\bf P2}. 

In any case, thus we believe that ~~{\bf P3} ~~is stronger than ~~{\bf P2} ~~and also the Brands-Chaum protocol since it does not require a previous set up to enable secure commit.

\paragraph*{Dishonest prover, closer attacker.} ~~{\bf P3} ~~is vulnerable with partial collusion when ~~$i$ ~~responds with Msg~2 and forwards ~~$n_v$, ~~and ~~$n_i$ ~~to ~~$p$ ~~later so that it can send the signature in Msg~5 to ~~$v$ ~~with ~~$n_v$, ~~$n_i$, ~~and other elements. However, ~~$p$ ~~does not share any secrets with ~~$i$ ~~to enable this attack. Hence, this protocol seems weaker than Brands-Chaum in this aspect.

\paragraph*{Farther attacker.} ~~{\bf P3} ~~is also vulnerable to the ~~``nearest-neighbor" ~~attack that ~~{\bf P2} ~~was, if we assume verifier does not know who it is talking to before receiving the signature in the final message. However, it would be unreasonable to make this assumption since the prover identity is explicitly included in the prior messages. Hence, we instantiated the prover variable ~~$P$ ~~to a ground atomic value in the verifier strand when we tested this protocol, whence we could not reproduce the ``nearest-neighbor" attack.

\paragraph*{Tweaking P3.} ~~Since the protocol was resistant to all other scenarios except collusion, we tweaked with the protocol to appreciate the significance of individual elements and their placement in messages. We could not find the use or purpose of the field $\mi{POS_P}$ described 						anywhere in \cite{MPPCS07} but removing it did not reveal any new attack. It is interesting to ask if the nonce ~~$N_P$ ~~inside Msg~4.2 is necessary. Removing it 					revealed an attack (Fig. \ref{f.P3-Protocol-Attack}.b).

\subsection{P4 - Guttman et al. \cite{GHST08}}

\begin{center}
\begin{figure}[h]
\scalebox{1.0}
{\includegraphics*[viewport=70 550 540 730]{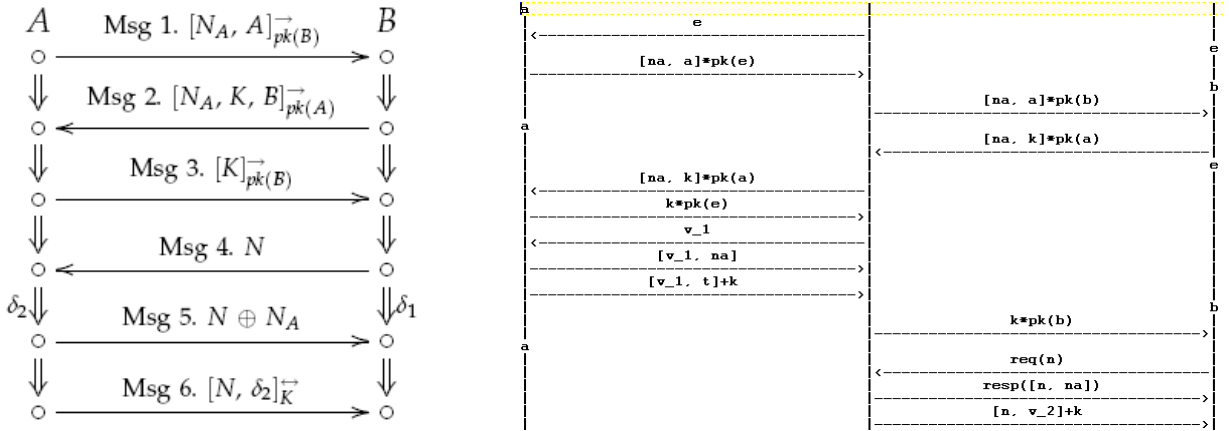}}
\caption{~~(a)~~Guttman et al. protocol {\bf P4} ~~~~~~~~(b) ~~Screen-shot of attack trace from the solver on the new Guttman et al.'s protocol. {\bf Note}: ~~v\_1, ~~v\_2 are variables;  ~~{\tt [t]*k} ~~= ~~$[t]^{\to}_k$, ~~~~~~~~~~~~~~~~{\tt [t]+k} ~~= ~~$[t]^{\leftrightarrow}_k$.}\label{f.P4-Protocol-Attack}
\end{figure}
\end{center}

{\bf P4} differs from all others in having more than one encrypted message in the authentication part, seemingly extending the NSPK/NSL protocols (Fig. \ref{f.P4-Protocol-Attack}.a). 

We analyzed this protocol with one strand per role in Scenarios~A and 1; i.e. we considered an honest verifier ~~$B$ ~~and an honest prover ~~$A$ ~~with a MITM attacker who is physically in between them. Further, as usual, we tied the ~~$\mi{Request}$ (Msg~4) ~~and ~~$\mi{Response}$ (Msg~5) ~~together in the node interleaving. Without ~~`$B$' ~~in Msg~2, the solver reported the trace with a MITM attack termed ``Lowe style" attack in \cite{GHST08} (Fig. \ref{f.P4-Protocol-Attack}.b).

In the trace, the attacker plays MITM between ~~$a$ ~~and ~~$b$ ~~and learns ~~$k$. ~~Then, the ~~$\mi{Response}$ ~~$[n,~n_a]$ ~~is sent from the attacker's location, which is physically closer to the verifier ~~$b$, ~~violating SDB and also follows it up with an authentication of the challenge ~~$n$ ~~in the last message ~~($[n,~\delta_2]^{\leftrightarrow}_k$).

The crux of this attack is the attacker's ability to satisfy both the conditions in Theorem~\ref{t.closer}. Satisfying the DB Part is trivial, but satisfying the authentication part is possible only by breaking the secrecy of ~~$k$ ~~since it is required to construct the last message, ~~$[n,~\delta_2]^{\leftrightarrow}_k$.

With larger semi-bundles/runs, more attacks could be possible by failing authentication even after the inclusion of ~~`$B$' ~~in Msg~2; E.g., see attacks on NSL given in \cite{MS01}.

\section{Conclusion}\label{s.conclusion}

In this paper, we described a method to automatically analyze distance bounding protocols. We formalized the main property of secure distance bounding and explained how violations of it can be tested using the constraint solver. We also illustrated our technique by presenting analyses of some published protocols.

A natural extension to our work is to extend it to unbounded analysis since the constraint solver only considers bounded number of protocol processes. Unbounded verification tools such as ~~{\sf ProVerif} ~~could be extended by tying the {\em Request} and {\em Rapid Response} together in the node interleavings as explained in Section~\ref{s.testing}, to produce attacks or to prove the absence of. In the case of ~~{\sf ProVerif}, ~~this is as simple as adding four events in the protocol, two each for the verifier and prover in the protocol, corresponding to sending and receiving the {\em Request} and {\em Rapid Response} respectively. No other change in the tool is required.

%An interesting question to ask is whether we considered all possible scenarios for attacks in Section~\ref{s.scenarios}: We have included all types of scenarios imaginable and reported in the literature, but it is hard to be sure that there are no others.

Other areas for future work include extending our framework with multiple penetrators in the 3D space, analyzing other properties in this model such as denial of service, obtaining decidability results for distance bounding, and testing protocols with a more powerful solver that considers message operators with algebraic properties such as Exclusive-OR. 

\paragraph*{Recent related work.} While the work in this paper was in progress, a related approach to verifying DB protocols using Isabelle/HOL was also in progress and is about to appear in \cite{SSBC09}. Being a verification effort, that approach differs 					from ours in the classical way that model checkers 					differ from theorem provers: the former tests for 					attacks while the latter proves the absence of. 						However, our approach can also be extended easily 					to unbounded verification with ~~{\sf ProVerif}, ~~as explained above. ~~{\sf ProVerif} ~~usually verifies protocols in a fraction of a second, faster than most theorem provers. But to be fair to the authors of \cite{SSBC09}, they consider other protocols used in wireless networks, not merely distance bounding as we did. In that sense, their work can be considered more elaborate than ours.

%\renewcommand{\baselinestretch}{0.98}
%\bibliographystyle{apalike}
%{\small
%\bibliography{SECRYPT09bib}}
%\renewcommand{\baselinestretch}{1}

{\small 
\bibliographystyle{plain}
\bibliography{FSTTCS09b}
}

\newpage
\appendix

\section{Constraint solving \cite{MS01}}\label{sa.consolv}

\subsection{Reduction Procedure \textbf{P}}

\begin{center}
 \begin{tabbing}
  \= ~~~$C :=$~initial constraint sequence \\
  \> ~~~$\sigma := 0$ \\
  \> ~~~\texttt{repeat} \\
  \> ~~~~~~ \texttt{let} $c^* = m : T$ be \= the constraint in $C$ \\
  \> \>  s.t. $m$ is not a variable \\
  \> ~~~~~~ \texttt{if} $c^*$ \= not found \\
  \> \> \texttt{output} \textbf{Satisfiable!} \\
  \> ~~~~~~apply rule \textit{(elim)} to $c^*$ until no longer applicable \\
  \> ~~~~~~$\forall r \in R$ \\
  \> ~~~~~~~~~\texttt{if} $r$ \= is applicable to $C$ \\
  \> \> $\langle C';\sigma'\rangle := r(C;\sigma)$ \\
  \> \> create node with $C'$; add $C \to C'$ edge \\
  \> \> push $\langle C'; \sigma' \rangle$ \\
  \> ~~~~~~~~~$\langle C;\sigma\rangle :=$ \texttt{pop} \\
  \> ~~~\texttt{until emptystack}  
 \end{tabbing}
 Reduction Procedure \textbf{P}~\cite{MS01}
\end{center}

\subsection{Set of reduction rules, $R$}

%\begin{center}
       \[ \frac{C_<,~m : T,~C_>;~\sigma}{\tau C_<,~\tau C_>;~\tau~\cup~\sigma }~~\mathrm{where}~\tau = \mathrm{mgu}(m,t) \wedge t \in T~~~(\mi{un})  \]
       \[ \frac{C_<,~[m_1,m_2]~:~T,~C_>;~\sigma }{C_<,~m_1~:~T,~m_2~:~T,~C_>;~\sigma }~~~(\mi{pair}) \]
       \[ \frac{ C_<,~h(m)~:~T,~C_>;~\sigma}{C_<,~m~:~T,~C_>;~\sigma}~~~( \mi{hash} ) \]
       \[ \frac{C_<,~[m]^{\to}_k~:~T,~C_>;~\sigma }{ C_<,~k~:~T,~m~:~T,~C_>;~\sigma }~~~(\mi{penc}) \]
       \[ \frac{C_<,~[m]^{\leftrightarrow}_k~:~T,~C_>;~\sigma }{C_<,~k~:~T,~m~:~T,~C_>;~\sigma }~~~(\mi{senc}) \]
       \[ \frac{C_<,~\mathrm{sig}_{\mi{pk}(\epsilon)}(m)~:~T,~C_>;~\sigma}{C_<,~m~:~T,~C_>;~\sigma}~~~(\mi{sig}) \]
       \[ \frac{C_<,~m~:~[t_1,t_2]~\cup~T,~C_>;~\sigma}{C_<,~m~:~t_1~\cup~t_2~\cup~T,~C_>;~\sigma }~~~(\mi{split}) \]
       \[ \frac{ C_<,~m~:~[t]^{\to}_{\mi{pk}(\epsilon)}~\cup~T,~C_>;~\sigma }{ C_<,~m~:~t~\cup~T,~C_>;~\sigma  }~~~(\mi{pdec}) \]
       \[ \frac{ C_<,~m~:~[t]^{\to}_k~\cup~T,~C_>;~\sigma  }{ \tau~C_<,~\tau~m~:~\tau~[t]^{\to}_k~\cup~\tau~T,~\tau~C_>;~\tau~\cup~\sigma  }, \]
        $\mathrm{where}~\tau = \mathrm{mgu}(k,\mi{pk}(\epsilon)), k \neq \mi{pk}(\epsilon)~~~(\mi{ksub})$ \\
       \[ \frac{C_<,~m~:~[t]^{\leftrightarrow}_k~\cup~T,~C_>;~\sigma }{ C_<,~k~:~T,~m~:~T~\cup~t~\cup~k,~C_>;~\sigma  }~~~(\mi{sdec}) \]
%\end{center}         

\subsection{Algorithm PB}\label{ss.algorithmPB}

We describe a simple extension to {\bf P} to generate bundles from satisfiable constraint sequences which in turn were generated from strands in a semi-bundle. Further, a single penetrator strand with the number of nodes equal to the sum of the nodes in the semi-bundle that captures all the reduction rules applied to generate terms on $-$ nodes or solve constraints. We name this new algorithm, {\bf PB}.

In {\bf PB}, we will assume a look-up table such as below to find out the weights corresponding to each attacker action. \\

\begin{center}
\begin{tabular}[!ht]{| l | l | l |}

\hline
{\bf Action} & {\bf Parameters} & {\bf Time taken} \\ 
\hline
$\mathsf{pair}$ & $m_1, m_2$ & ($m_1$.len + $m_2$.len) \\
\hline
$\mathsf{split}$ & $[m_1, m_2]$ & $(m_1 + m_2)$.len \\
\hline
$\mathsf{senc, sdec}$ & $\{ m, k \}$ &  $m.{\rm len} \times k.{\rm len}$ \\
\hline
$\mathsf{penc, pdec}$ & $\{ m, k \}$ &  $\mathrm{1000} \times m.{\rm len} \times k.{\rm len}$ \\
\hline
$\mathsf{hash}$ & $m$ &  $m.{\rm len} \times 10$ \\
\hline
$\mathsf{sig}$ & $\{ m, k \}$ &  $\mathrm{1000} \times m.{\rm len} \times k.{\rm len}$ \\ \hline 
\end{tabular}
\end{center}

\vspace{0.2in}

{\bf Note:} $m$.len denotes the number of bytes in $m$. \\

As is obvious from the above table,
we adopt the well known fact that asymmetric key encryption /decryption is about thousand times slower than its symmetric counterpart. Note that rules $\mathsf{un}$ and $\mathsf{ksub}$ do not correspond to any penetrator action but only generate the substitution required to complete the semi-bundle to a bundle.

The  time of traversal for a message depends on the message length, distance and the velocity of the signal. We assume that there is an appropriate formula for an environment to calculate that time, using those parameters. \\

%The calculation of time is a recursive process. i.e., the time taken to construct a message is the total of times taken to construct all its (proper) subterms. This can be calculated for a term by ``solving" the constraint on which the term is a target although in the algorithm below, we focus on calculating all the steps needed to be taken by an attacker to generate a term using a given set of terms which includes decrypting known encryptions and splitting pairs. This captures all the actions taken by an attacker/prover to generate a message for the verifier with the existing knowledge at that point. \\

\noindent
{\hrule height 2pt} 
\vspace{0in}
		\begin{flushleft} 
						{\bf Algorithm} \emph{ProduceBundle}				 				
		\end{flushleft}
\vspace{0in}
{\hrule height 1pt}

\begin{tabbing}
{\bf Input:} Semi-bundle $S$, Constraint solving procedure {\bf P} \\
{\bf Output:} Bundle. \\[2pt]
\= 1 \= {\em Draw} all the strands in $S$ (with $\Rightarrow$ edges)\\
\> 2 \> {\em Label} nodes on each strand with (time, sign, term) \\
\> 3 \> {\bf fo}\={\bf r each} node merge $\mathcal{N}$ from $S$\\
\> 4 \> \> {\em Generate} a constraint sequence $C$ and {\em solve} $C$ with {\bf P}\\
\> 5 \> \> {\bf if} $C$ not satisfiable, {\bf continue;}\\
\> 6 \> \> {\bf fo}\={\bf r each} successive node $m$ in $\mathcal{N}$ \\
\> 7  \> \> \> {\em draw} a penetrator node $n$ connecting it to the \\
\> 8 \> \> \> previous node on the same strand using a $\Rightarrow$;\\
\> 9 \> \> \> {\bf if} \= sign($m$)=`$+$' {\bf then}\\
\> 10 \> \> \> \> {\em draw} an edge $\to$ between $m,n$; \\
\> 11 \> \> \> \> {\em update} time on $n$ as time($m$) + weight($m,n$); \\
\> 12 \> \> \> {\bf if} \= sign($m$)=`$-$' {\bf then}\\
\> 13 \> \> \> \> mark the weight of $\Rightarrow$ edge as the sum of weights \\
\> 14 \> \> \> \> of all the rules applied to satisfy the constraint; \\
\> 15 \> \> \> \> {\em draw} an edge $\to$ between $n, m$; \\
\> 16 \> \> \> \> {\em update} time on $m$ as time($n$) + weight($n, m$); 
\end{tabbing}

{\hrule height 1pt}

\vspace{0.2in}

%In {\bf PB} line~14 --- ``all the rules applied to satisfy the constraint", can be interpreted to mean only those rules that are really needed to construct $m$ in $m~:~T$ and that can be fired only after receiving a term. This can optimize time to respond, make the attacker stronger and enable additional attacks. For example, consider $[n_a,n_b]_k : T \cup \{ k \}$ with $\{ n_a, n_b \} \in T$ and $k$ to be received before solving the constraint; $\{n_a, n_b\}$ can be paired as $[n_a,n_b]$ before receiving $k$ and hence the cost of the $\msf{pair}$ rule need not be added to the cost of penetrator actions to produce $[n_a, n_b]_{k}$.

%Not all constraints are ordered (i.e., nodes to which the left side of constraints belong to need not have an order since the relation $\prec$ is not defined between nodes on different semi-strands). This concept (the possibility of ``jumping the gun" on sending messages) can be incorporated when calculating weights, by redacting the weights of all the unnecessary edges\footnote{Edges corresponding to verifier requests are not necessary for prover's responses that follow if there are no common subterms in the request and response.} in the path.

%%%%%%%%%%%%%%%%%%%%%%%%%%%%%%%%%%%%%%%%%%%%%%%%%%%%%%%%%%%%
%=========== 	EXAMPLE - NSPK ===============================
%%%%%%%%%%%%%%%%%%%%%%%%%%%%%%%%%%%%%%%%%%%%%%%%%%%%%%%%%%%%

\subsection{Algorithm PB - An example}\label{sa.alg-example}

Consider the Needham-Schroeder Public-Key (NSPK) protocol \cite{NS78}:

\vspace{-0.3in}
\begin{displaymath}\label{f.NSPK}
		\xymatrix@=0pt@C=60pt@R=12pt@W=0pt{		
    		\save[]+<0cm,-0.3cm>*{A} \restore & \save[]+<0cm,-0.3cm>*{B} \restore \\
     	 {11} \ar[r]^{[N_A, A]_{\mi{pk}(B)}}\ar@2{->}[d]   &   {21}\ar@2{->}[d]  \\
       {12}\ar@2{->}[d]    &     \ar[l]_{[N_A, N_B]_{\mi{pk}(A)}}  {22}\ar@2{->}[d]   \\
     	 {13} \ar[r]^{[N_B]_{\mi{pk}(B)}}   &     {23}
         } 
\end{displaymath}

Following our procedure {\bf PB}, we first draw the semi-strands $a$ for $A$ and $b$ for $B$. We then  consider the node merge $\langle 11, 21, 22,$ $12, 13, 23 \rangle$ and from it the constraint sequence, $\langle 21~:~11, 12~:~\{ 11, 22 \},$ $23~:~\{11, 22, 13 \} \rangle$. This sequence will reveal Lowe's attack on NSPK \cite{LoweBF96}.

Following our algorithm, 

\begin{enumerate}
	\item we first add a penetrator node $e1$ and to it a $\to$ 					edge from $11$, 
	\item send $[n_a, a]_{\mi{pk}(e)}$ with it's weight as the 						product of the distance between $a$ and the penetrator 					$e$, the length of the message $[n_a, a]_{\mi{pk}(e)}$ 					and the velocity of the signal, 
	\item update the time on $e1$ with that weight counting the 					time on node $11$ as zero, 
	\item add a $\Rightarrow$ edge from $e1$ to a second 									penetrator node $e2$, it's weight  $\delta_1$ being the 				sum of all the rules to solve the first constraint and 					generate the term on node $21$; i.e, $\delta_1$ = time 					taken to apply $\msf{pdec}$, $\msf{split}$, $\msf{pair}$ 				 and $\msf{penc}$ after appropriately parameterizing 						  with message and key lengths,
	\item update the node's time as the weights of edges 									$11 \to e1$ and $e1 \Rightarrow e2$ and add a $\to$ edge 				 from $e2$ to $21$. 
\end{enumerate}
	
	Similarly, we can finish the other nodes following their order in the node merge:
	% (with $\delta_3 = 0$, $\delta_4$ equals weight of $(12,13)$ and $\delta_5$ = time taken to apply $\msf{pdec}$ on $[n_b]_{\mi{pk}(\epsilon)}$ and $\msf{penc}$ on $n_b$ and complete the bundle):
%($\delta_3 = 0$, $\delta_2, \delta_4$ are preset):

%  

\vspace{-0.3in}
\begin{displaymath}
		\xymatrix@=0pt@C=60pt@R=12pt@W=0pt{		
     		\save[]+<0cm,-0.3cm>*{a} \restore & \save[]+<0cm,-0.3cm>*{e} \restore & \save[]+<0cm,-0.3cm>*{b} \restore \\
     	 {11} \ar@2{->}[ddd]\ar[r]^{[n_a, a]_{\mi{pk}(e)}}   &  {e1}\ar@2{->}[d]^{\delta_1}    & {} \\
     	            {}       & {e2}\ar[r]^{[n_a, a]_{\mi{pk}(a)}}\ar@2{->}[d]^{\delta_2}  &   {21}\ar@2{->}[d]  \\
       								     & {e3}\ar@2{->}[d]^{\delta_3} &   \ar[l]_{[n_a, n_b]_{\mi{pk(a)}}}{22}\ar@2{->}[ddd]   \\
	            {12}\ar@2{->}[d]    & \ar[l]_{[n_a, n_b]_{\mi{pk(a)}}}{e4}\ar@2{->}[d]^{\delta_4}     &   {}   \\
     	 {13} \ar[r]^{[n_b]_{\mi{pk}(e)}}   &  {e5}\ar@2{->}[d]^{\delta_5}    & {} \\
     	            {}       & {e6}\ar[r]^{[n_b]_{\mi{pk}(b)}}  &   {23}  \\
         } 
\end{displaymath}

\section{Proofs}\label{sa.proofs}

\begin{theorem}~~{\em\bf [Termination, Soundness and Completeness]}\label{ta.termination}

	Algorithm {\bf PB} terminates and is sound and complete.

\end{theorem}

\begin{proof}
	
	Consider the steps in the algorithm {\bf PB} sequentially:
	
	\begin{enumerate}
	
		\item {\bf PB} draws finitely many strands each with finitely many nodes 							from a given semi-bundle $S$; 
	
		\item	Next, it generates finitely many node merges $\mathcal{N}$ and solves 						each of them using {\bf P} which is proven to be terminating, sound and 					complete \cite{MS01};
	
		\item Finally, it loops for all the finitely many  nodes in 					$\mathcal{N}$. In each iteration, it performs all 							atomic actions, namely drawing a node or an edge and 						updating times on node labels; The only non-atomic 							action is the adding of  weights of all the finitely 						many reduction rules to solve a constraint.
		
\end{enumerate}

Since all the above are finitely many actions, {\bf PB} terminates. Further, its soundness and completeness follow directly from those properties of {\bf P} and since every node in the semi-bundle is handled. 

\end{proof}

As promised in Section~\ref{ss.loc-scenarios}, below we prove that in Scenario~1, if an attacker can generate all the messages expected by the verifier from the $\mi{Request}$ to the $\mi{Response}$ without those messages being sent by the prover, then SDB is violated if all other messages expected by the verifier can also be generated by the attacker with or with out those messages emanating from the prover.

\begin{theorem}\label{ta.closer}

	Suppose $t_0, t_1, \ldots, t_m$ are terms on $m$ nodes on the  verifier strand $v$ with time of flight measured in between $t_0$ and $t_m$. Then, there exists a bundle with a violation of SDB if 

\begin{itemize}
		\item the constraints $\langle t_1 : T_0, \ldots ,t_m : T_m 					\rangle$ are satisfiable where for $i = 0$ to $m$, 							every $t \in T_i$ either belongs to $T_0$ or a $+$ 							node on $v$ and every $t_i$ is a term on a $-$ node on 					 $v$;
		\item all other $-$ nodes in $v$ are realizable.	
\end{itemize}

\end{theorem}

\begin{proof}
	Let the ideal bundle in between $v$ and a prover $p$ be denoted $B$. Let the distance between $v$ and $p$ be $d(v,p)$. Let the distance between $v$ and $i$ be $d(v,i)$ and let $d(v,i) < d(v,p)$; i.e., any edge between a node on $v$ to a node on $i$ will have lesser weight than the edge from $v$ to $p$ when the same number of bits are transmitted on the edges.
	
	Say the path between $t_0$ and $t_m$ is $w$ in $B$. Now consider a real bundle $B'$ where every $+$ node on $v$ between $t_0$ and $t_m$ is connected to a $-$ node on $i$ and vice-versa (which is possible since every $-$ node on $v$ is satisfiable without terms from $p$). Since the weights of $\pm \Rightarrow +$ are preset in the strands $v$ and $p$, the weights of those edges will be equal in both $i$ and $p$. Hence, the weight of the path between $t_0$ and $t_m$ in $B'$ will be lesser since the weight of the only remaining $\to$ edges are lesser as explained above.
	
	Hence, by Def.~\ref{d.SDB}, there is a violation of SDB in $B'$.
	
\end{proof}

We also prove that we can find attacks in Scenario~2 of Section~\ref{ss.loc-scenarios},  by analyzing protocols for potential executions with $p$ sending all the messages in the DB part and attacker sending the remaining messages, as promised in Section~\ref{ss.loc-scenarios}.

\begin{theorem}\label{ta.farther}

	Consider $v, p_1, p_2$ where $d(v,p_1) < d(v,p_2)$. Let $\langle t_0,\ldots,t_m \rangle$ be nodes on $v$ between which time of flight is measured.
	Then, there is a violation of SDB if 
	
	\begin{itemize}
		\item the constraints $\langle t_1~:~T_1, \ldots, t_m~:~T_m 					\rangle$ are satisfiable where for all $i = 1$ to $m$, 					 every $t_i$ is unified with some $t'_1 \in T_i$ where 						$t'_i$ is a term on $p_1$.
		\item All other `$-$' nodes of $v$ are realizable  						without unifying with any subterms of $p_1$.
	\end{itemize}
	
\end{theorem}

\begin{proof}
	Consider the ideal bundle to be between $v$ and $p_2$ denoted as $B$ and let the time of flight in $B$ be $w$.
	
	Now consider another bundle $B'$ produced by {\bf PB} wherein every $m \to n$ edge is such that $m$ has one of $t_0, \ldots, t_m$ as a term and $n$ is a node on $p_1$.
	
	This is possible since every constraint $\langle t_1~:~T_1, \ldots, t_m~:~T_m \rangle$ is satisfiable by unifying with a term in $p_1$ resulting in the attacker edge $\Rightarrow$ having a weight $0$. In this situation, an equivalent bundle can be produced where the attacker action $m \to e_i \Rightarrow e_j \to n$ (for some nodes $e_i$ and $e_j$ on the attacker strand) is replaced with a straight edge $m \to n$. Since $d(v,p_1) < d(v,p_2)$, the sum of weights of those edges in $B'$ will be lesser than $B$. Further, the weights of $\pm \Rightarrow +$ edges for $p_1$ in $B$ or $p_2$ in $B'$ will be equal since they are preset and constant from assumptions in the protocol model. 
	
	Thus, by Def.~\ref{d.SDB}, there is an attack on SDB in $B'$.
	
\end{proof}

We supplement these results with some general results on collusion. These results will show that collusion is in general impossible to prevent. However, while in some cases it works without any shared secrets between the prover and attacker, in some other cases it necessarily requires at least some shared secrets.

\begin{corollary}~~{\em\bf [Full collusion]}\label{c.full-collusion}

	It can be easily seen from the proof of Theorems~\ref{ta.closer} and \ref{ta.farther} that if $p$ colludes with $i$ and shares all its secrets with $i$, then it results in a direct violation of SDB since $i$ can send all the messages expected by $v$ without any involvement from $p$; i.e., all the constraints from $v$'s strand are satisfiable without the prover strand $p$ at all and due to $i$'s closer proximity, the weight of the path from the {\it Request} to {\it Response} used to calculate ToF will be lesser.
	
\end{corollary}

\begin{corollary}~~{\em\bf [Partial Collusion]}\label{c.partial-collusion}

	This collusion attack can work even under partial collusion; i.e., if $p$ initially shares with $i$ only those subterms that are needed to satisfy all the constraints between {\it Request} and {\it Response}, then from the proofs it can be easily seen that this is sufficient for an attack on SDB by simply relaying all the other messages from $p$ to $v$ that are not used to calculate the ToF.

\end{corollary}

\begin{corollary}~~{\em\bf [Preventing collusion]}\label{c.prevent-collusion}

	A simple way to prevent collusion is by measuring time of flight between encrypted messages, some of which must be decrypted with a private key known only to the legitimate but dishonest and colluding prover. In such a situation, violation of SDB is possible only when the prover shares its private keys or secrets with the attacker so that all the constraints in the DB part are satisfiable without the nodes from the prover strand. 
	
	There might be situations where provers collude with others, but do not want to share all their information, especially long-term keys or private keys. In those cases, collusion attacks can be prevented by measuring ToF between encrypted messages as explained above.
	
\end{corollary}	

\begin{corollary}~~{\em\bf [Collusion possible in both scenarios]}\label{c.all-collusion}

	These observations are true whether $i$ is close to $v$ than $p$ or otherwise. In the case of $i$ being further from $v$ than $p$, under collusion, $p$ attempts to deliberately make $i$ appear closer.
	
\end{corollary}

\section{Soundness and Completeness of implementation}\label{sa.sound-complete}

We will now prove soundness and completeness of our implementation wrt Def. \ref{d.SDB}. We make two assumptions about DB protocols for these results:

\begin{assumption}\label{a.consecutive}
	$\mi{Request}$ and $\mi{Rapid Response}$ are the only two 						consecutive nodes between which ToF is measured.
\end{assumption}

\begin{assumption}\label{a.lightest-edge}
	The elapsed time between the prover receiving the $\mi{Request}$ and sending the 
	$\mi{Rapid Response}$ is lower than the weight of $\Rightarrow +$ edges in the verifier or the prover strands.
\end{assumption}

DB protocols for networks with stringent time and resource constraints such as sensor networks, must satisfy both these assumptions \cite{BC95,MPPCS07}.

\subsection{Soundness}\label{ss.soundness}

The implementation can be considered sound wrt Def. \ref{d.SDB} if all the real bundles produced  by the tool necessarily violate SDB. In other words, every bundle that is produced should have the ToF lesser than the ideal bundle.

We made a minor change to the original constraint solver by Millen-Shmatikov to achieve soundness: we restricted it to consider only those node interleavings wherein the ~~{\em Request} ~~and ~~{\em Rapid Response} ~~nodes in the verifier strand immediately follow each other. We then tested protocols by following the results of Theorems \ref{t.closer} and \ref{ta.farther} in Section \ref{s.scenarios}.

To see why we restricted the interleavings, consider the following two bundles output by the solver for the extended-Echo protocol {\bf P1} introduced in Section~\ref{s.intro} (below $m = [n_v, ~v, ~p]$):

\vspace{-0.3in}
\begin{displaymath}\label{f.P1-Attack1}
		\xymatrix@1@-10pt@C=38pt@R=12pt@W=0pt{		
    		\save[]+<0cm,-0.3cm>*{v} \restore & \save[]+<0.3cm,-0.3cm>*{i(p), i(v)} \restore & \save[]+<0cm,-1.8cm>*{p} \restore & \save[]+<0cm,-0.3cm>*{v} \restore & \save[]+<0.3cm,-0.3cm>*{i(p), i(v)} \restore & \save[]+<0cm,-1.0cm>*{p} \restore  \\
{\circ} \ar[r]^-{n_v}\ar@2{->}[d]_{\delta_1}  &  {\circ}\ar@2{->}[d]_{\delta_2}  & {} & {\circ}\ar[r]^-{n_v}\ar@2{->}[dddd]_{\delta_1} & {\circ}\ar@2{->}[d] & {} \\
{\circ} \ar@2{->}[ddd]  &  \ar[l]_-{n_v}{\circ}\ar@2{->}[d] & {} & 
 {} & {\circ}\ar[r]^-{n_v}\ar@2{->}[d]_{\delta_2} & {\circ}\ar@2{->}[d]\\
 {}   &  {\circ}\ar@2{->}[d]\ar[r]^-{n_v}   &  {\circ}\ar@2{->}[d] & 
 {} & {\circ}\ar@2{->}[d] & \ar[l]_-{n_v}{\circ}\ar@2{->}[d] \\    
 {}   &  {\circ}   &  \ar[l]_-{n_v}{\circ}\ar@2{->}[d]  &  
 {} & {\circ}\ar@2{->}[d] & {\circ}\ar[l]_-{\mi{Sig}_{\mi{pk}(p)}(m)} \\
 {\circ}  & 	{}		&  	\ar[ll]_-{\mi{Sig}_{\mi{pk}(p)}(m)}{\circ} & {\circ}\ar@2{->}[d]  &  \ar[l]_-{n_v}{\circ}\ar@2{->}[d]   &  {}\\								
 {}  &  {}   &  {}  &  {\circ}  &  {\circ}\ar[l]_-{\mi{Sig}_{\mi{pk}(p)}(m)}   &   {}  \\
      } 
\end{displaymath}

These bundles were produced in an honest prover scenario with attacker closer to the verifier. Now the conditions of Theorem~\ref{ta.closer} are satisfied in both the bundles; i.e., the messages in the DB part are satisfiable without the send node in the prover strand to send the ~~{\em Rapid  Response}, ~~and the signature is realizable as well. Yet, only the bundle on the left obviously has a violation of SDB since its ToF ~(~$\delta_1$) ~~is lesser compared to the ideal case. On the other hand, the one on the right has its ~~$\delta_1$ ~~larger than the ideal case.  

The reason is that the bundle on the right was found on an interleaving where ~~{\em Request} ~~and ~~{\em Rapid Response} ~~do not immediately follow each other. In this case,  output of algorithm ~~{\bf PB} ~~produces a path ~~({\em Request},~~{\em Rapid Response}) ~~that is heavier than  a path between the same nodes in the ideal bundle, since it has to draw the edges corresponding to other realizable nodes before it draws the edges for the ~~{\em Rapid Response} ~~node.

Theorem~\ref{ta.closer} merely states that an attack bundle exists but does not point to it precisely. Restricting interleavings in this fashion restricts the solver so that every bundle it produces is necessarily an attack bundle, thereby achieving soundness.

\subsection{Completeness}\label{ss.completeness}

By restricting interleavings do we lose any attacks? We can easily prove that we do not. 

Firstly, let us note that every bundle produced by the solver after restricting interleavings is similar to the bundle on the left given in Section \ref{ss.soundness}. i.e., bundles where the attacker receives the ~~{\em Request} ~~and immediately sends the ~~{\em Rapid Response} ~~back to the verifier, without sending or receiving any other messages in between. To prove completeness, we will show that these are the ~~{\em only} ~~bundles where there is a violation of SDB.

To achieve a contradiction, consider the bundle below, which is not produced by the solver when we restrict interleavings (the attacker does not send the ~~{\em Rapid Response} ~~immediately after receiving the ~~{\em Request}):

\begin{displaymath}\label{f.real1}
		\xymatrix@=0pt@C=30pt@R=12pt@W=0pt{		
				\save[]+<0cm,-0.3cm>*{v} \restore & 														\save[]+<0.3cm,-0.3cm>*{p'} \restore & 									\save[]+<0cm,-1.0cm>*{p} \restore  \\
				*+[o][F-]{1}\ar[r]^-{}\ar@2{->}[ddd]_{} & 																		*+[o][F-]{2}\ar@2{->}[d] & {} \\
 				{} & *+[o][F-]{3}\ar[r]^-{}\ar@2{->}[d]_{} & 																*+[o][F-]{4}\ar@2{->}[d]\\
 				{} & *+[o][F-]{6}\ar@2{->}[d] & 																							*+[o][F-]{5} \ar[l]^{} \\  
				*+[o][F-]{8}  &  *+[o][F-]{7} \ar[l]^{}  &  {}\\								
  } 
\end{displaymath}

Call this bundle ~~$B$. ~~Say there is a violation of SDB in ~~$B$ ~~when the ideal bundle is between ~~$v$ ~~and ~~$p$. ~~That means, the weight of the path ~~$(2,~7)$ ~~is lower than the weight of edge ~~$(2,~3)$ ~~in the ideal bundle ~~$B'$ ~~below between ~~$v$ ~~and ~~$p$ ~~(note that the weight of ~~$\to$ ~~edges in the path ~~$(1,~8)$ ~~in ~~$B$ ~~is equal to the weight of ~~$\to$ ~~edges between ~~$(1,~4)$ ~~in ~~$B'$): 

\begin{displaymath}\label{f.ideal}
%	\entrymodifiers={++[o][F-]}
%	\SelectTips{cm}{}
		\xymatrix@=0pt@C=70pt@R=12pt@W=0pt{		
   		\save[]+<0cm,-6cm>{v} \restore & \save[]+<0cm,-6cm>{p} \restore \\
     	 *+[o][F-]{1} \ar[r]^{}\ar@2{->}[d] & 																								*+[o][F-]{2}\ar@2{->}[d]  \\
       *+[o][F-]{4} &   *+[o][F-]{3} \ar[l]^{}   \\
    } 
\end{displaymath}

Now the weight of ~~$(2,~3) + (3,~6) + (6,~7)$ ~~in ~~$B$ ~~is obviously more than the weight of edge ~~$(4,~5)$ ~~in ~~$B$. ~~From above, since there is an attack in ~~$B$, ~~the weight of ~~$(2,~3) + (3,~6) + (6,~7)$ ~~must be lesser than ~~$(2,~3)$ ~~in ~~$B'$. ~~However, the weight of ~~$(4,~5)$ ~~in ~~$B$ ~~itself is greater than or equal to the weight of ~~$(2,~3)$ ~~in ~~$B'$ ~~from Assumption \ref{a.lightest-edge}, a contradiction. 

If ~~$B$ ~~depicts a violation of SDB in the farther attacker scenario (i.e., ~~$p'$ ~~is a prover strand), then by Assumption \ref{a.lightest-edge}, the weight of ~~$(2,~3) + (3,~6) + (6,~7)$ ~~in ~~$B$ ~~will be heavier than ~~$(2,~3)$ ~~in ~~$B'$, ~~again a contradiction.

Finally, if there is a bundle with ToF lighter than ~~$(1,4)$ ~~in ~~$B'$ ~~that is not produced by the solver, ~~then it must be from a node interleaving with a maximum of one send or receive node per strand in between ~~{\em Request} ~~and ~~{\em Rapid Response}. ~~For instance, bundle ~~$B{''}$ ~~below:

\begin{displaymath}\label{f.real2}
		\xymatrix@=0pt@C=30pt@R=12pt@W=0pt{		
				\save[]+<0cm,-0.3cm>*{v} \restore & 														\save[]+<0.3cm,-0.3cm>*{p'} \restore & 									\save[]+<0cm,-1.0cm>*{p} \restore  \\
				*+[o][F-]{1}\ar[r]^-{}\ar@2{->}[dd]_{} & 																		*+[o][F-]{2}\ar@2{->}[d] & {} \\
 				{} & *+[o][F-]{3}\ar[r]^-{}\ar@2{->}[d]_{} & 																*+[o][F-]{4} \\
				*+[o][F-]{8}  &  *+[o][F-]{7} \ar[l]^{}  &  {}\\								
  } 
\end{displaymath}

However, an attack on bundles such as these also implies an attack on a bundle where node ~~3 ~~occurs before node ~~2 ~~or after node ~~4, ~~which are indeed produced by the solver.

Thus, by restricting interleavings to just those with $(\mi{Request},\mi{Response})$ tied together, we do not lose any attack.

\section{More attacks}\label{sa.more-attacks}

\subsection{P2 - Brands \& Chaum}

\paragraph*{Farther attacker.} This protocol forms a nice example to test under Scenario~2. Assuming attacker is further away from the verifier, we followed our results in Section~\ref{s.scenarios} and removed the nodes in the DB part in one strand while removing the signature (Msg~5) in another strand. The solver then output the following attack simplified by removing some relays:

	\begin{displaymath}\label{f.P4-Attack1}
		\xymatrix@=0pt@C=75pt@R=12pt@W=0pt{		
    		\save[]+<0cm,-0.3cm>*{v} \restore & \save[]+<0.0cm,-0.3cm>*{p} \restore & \save[]+<0cm,-1.8cm>*{i} \restore \\
	  {\circ}\ar[r]^-{v}\ar@2{->}[d] & {\circ}\ar@2{->}[d]  & {}\\
	 							 {\circ}\ar@2{->}[d]  & \ar[l]_-{n_p}{\circ}\ar@2{->}[d] & {}\\
	   {\circ}\ar[r]^-{n_v}\ar@2{->}[d]_{\delta_1} & {\circ}\ar@2{->}[d]^{\delta_2}  & {}\\
				 {\circ}\ar@2{->}[dd]  & \ar[l]_-{n_v~\oplus~ n_p}{\circ}\ar@2{->}[d] & {}\\
		{} & {\circ}\ar[r]^-{\mi{Sig}_{\mi{pk}(p)}[n_v,~n_p]} &  {\circ}\ar@2{->}[d] \\
  {\circ} & {} & \ar[ll]_-{\mi{Sig}_{\mi{pk}(i)}[n_v,~n_p]}{\circ}\\
     } 
	\end{displaymath}
	
Here, ~~$i$ ~~is further away from ~~$v$ ~~than ~~$p$ ~~and possibly out of a range that ~~$v$ ~~wishes to include nodes. ~~$i$ ~~then lets ~~$p$ ~~respond to ~~$v$'s ~~request, obstruct its authenticated response (Msg 5) and substitutes its own message signed with its own private key. 

Obviously, this attack cannot work with the prover identity inside the signature, but only when the verifier uses the protocol to find its nearest neighbor.

\end{document}